\documentclass[
  aps,
  prl,
  reprint,
=  amsmath,amssymb,
  superscriptaddress,   
  footinbib,
]{revtex4-2}

\usepackage{graphicx}
\usepackage{siunitx}      
\usepackage{bm}           
\usepackage[version=4]{mhchem}
\usepackage{color}
\usepackage{soul}
\usepackage{siunitx}
\usepackage{hyperref}

\DeclareSIUnit{\rlu}{r.l.u.}

\hypersetup{
    colorlinks=true,
    linkcolor=black,    
    citecolor=black,    
    filecolor=black,    
    urlcolor=black      
}

\begin{document}

\title{Destructive Interference of a Charge Density Wave in the Skyrmion Magnet \ce{GdRu2Si2}}

\author{G. D. A. Wood}
\email{george.wood@stfc.ac.uk}
\affiliation{ISIS Neutron and Muon Source, STFC Rutherford Appleton Laboratory, Harwell Science and Innovation Campus, United Kingdom}
\affiliation{Department of Physics, University of Warwick, Coventry, CV4 7AL, United Kingdom}

\author{D. A. Mayoh}
\affiliation{Department of Physics, University of Warwick, Coventry, CV4 7AL, United Kingdom}

\author{F. Orlandi}
\affiliation{ISIS Neutron and Muon Source, STFC Rutherford Appleton Laboratory, Harwell Science and Innovation Campus, United Kingdom}

\author{A. Bombardi}
\affiliation{Diamond Light Source Ltd, Harwell Science and Innovation Campus, Didcot, Oxfordshire, United Kingdom}

\author{D. D. Khalyavin}
\affiliation{ISIS Neutron and Muon Source, STFC Rutherford Appleton Laboratory, Harwell Science and Innovation Campus, United Kingdom}

\author{P. Manuel}
\affiliation{ISIS Neutron and Muon Source, STFC Rutherford Appleton Laboratory, Harwell Science and Innovation Campus, United Kingdom}

\author{S. H. Moody}
\affiliation{Laboratory for Neutron Scattering and Imaging (LNS), PSI Center for Neutron and Muon Science, Paul Scherrer Institut (PSI), CH-5232 Villigen, Switzerland}
\affiliation{Department of Physics, Durham University, South Road, Durham DH1 3LE, United Kingdom}

\author{P. D. Hatton}
\affiliation{Department of Physics, Durham University, South Road, Durham DH1 3LE, United Kingdom}

\author{R. D. Johnson}
\affiliation{Department of Physics, Durham University, South Road, Durham DH1 3LE, United Kingdom}

\author{T. W. Chamberlain}
\affiliation{Department of Physics, University of Warwick, Coventry, CV4 7AL, United Kingdom}

\author{O. A. Petrenko}
\affiliation{Department of Physics, University of Warwick, Coventry, CV4 7AL, United Kingdom}

\author{M. R. Lees}
\affiliation{Department of Physics, University of Warwick, Coventry, CV4 7AL, United Kingdom}

\author{G. Balakrishnan}
\email{g.balakrishnan@warwick.ac.uk}
\affiliation{Department of Physics, University of Warwick, Coventry, CV4 7AL, United Kingdom}

\date{\today}

\begin{abstract}

In skyrmion magnets, charge density waves induce higher-order exchange interactions that can select multi-Q ground states. Using resonant X-ray scattering and neutron diffraction, we measure temperature-dependent spin-charge coupling in \ce{GdRu2Si2}. A minimal Landau model with one critical exponent and two ordering temperatures reproduces all six observed order parameters. Spin-density wave components of the multi-Q ground state drive charge ordering, but the onset of a magnetic helix suppresses it via destructive interference. 

 \end{abstract}

\maketitle

Coupling between magnetic and charge degrees of freedom is a key phenomenological aspect of several condensed matter systems, including colossal magnetoresistance materials~\cite{PhysRevLett.75.914,renner2002atomic}, high-$T_{c}$ superconductors~\cite{RevModPhys.87.457,PhysRevB.57.1422,PhysRevLett.101.257003}, spin-Peierls transitions~\cite{PhysRevLett.35.744,PhysRevB.19.402,PhysRevLett.70.3651} and multiferroics~\cite{hill2000there,kimura2003magnetic,PhysRevLett.108.067201}. At the most general level, when magnetic order couples to charge (or lattice) distortions, electronic interactions are periodically modified, leading to higher-order effective spin interactions and emergent behavior~\cite{PhysRev.120.335,PhysRevLett.88.067203}. This is cardinal in multi-Q magnets where higher-order exchange terms can select multi-Q order in centrosymmetric systems~\cite{PhysRevB.95.224424,PhysRevLett.118.147205,hayami2021topological,sharma2023machine,hayami2024stabilization,PhysRevB.103.064414}. Indeed, in intermetallic multi-Q magnets, where Ruderman-Kittel-Kasuya-Yosida (RKKY) interactions mediate exchange between local moments~\cite{kurumaji2019skyrmion,hirschberger2019skyrmion,khanh2020nanometric,takagi2022square,yoshimochi2024,rathnaweera2026antibonding}, it is clear that magnetic-charge coupling inseparably links spin textures with charge density wave (CDW) order~\cite{yang2025incommensurate,PhysRevB.104.144404,yasui2020imaging,spethmann2024sp,butler2026intra}.

\begin{figure}[h]
\includegraphics[width=0.8\linewidth,clip]{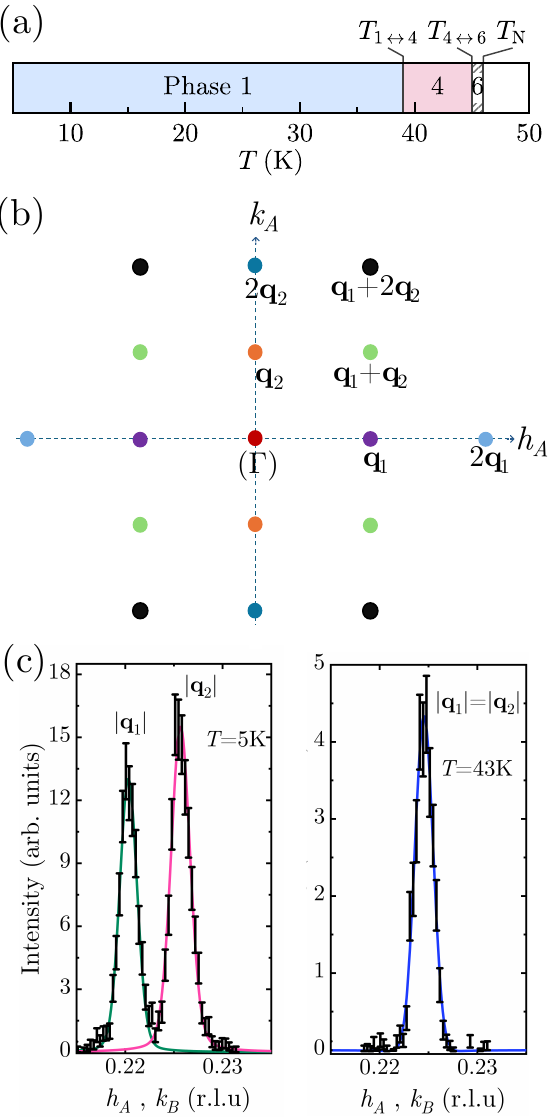}
\caption{\label{fig:1}(a) Zero-field phase diagram of \ce{GdRu2Si2}. Above $T_N$ \ce{GdRu2Si2} is paramagnetic. (b) Map of the incommensurate satellites observed in \ce{GdRu2Si2} due to magnetic and charge ordering. (c) Zero-field single-crystal neutron diffraction measurements of the length of the primary propagation vectors $\mathbf{q}_{1}$ and $\mathbf{q}_{2}$. Here, $h_{A}=(h,0,0)_{A}$ and $k_{B}=(k,0,0)_{B}$, given $A$ and $B$ domains are rotationally orthogonal, $h_{A}$ and $k_{B}$ denote equivalent directions in the instrument frame. In the ground state at \SI{5}{\kelvin}, these propagation vectors are distinct, but in Phase 4 at \SI{43}{\kelvin} they are equal in length within our experimental resolution.}
\end{figure}

One such intermetallic system is \ce{GdRu2Si2}, which hosts a square skyrmion lattice amongst several multi-Q magnetic structures~\cite{khanh2020nanometric,khanh2022zoology,PhysRevB.111.054440}, with concomitant CDW order~\cite{yasui2020imaging,spethmann2024sp}. Three magnetic phases exist in zero field, labeled Phase 1, 4, and 6 [see Fig.~\ref{fig:1} (a)]~\cite{gries2025uniaxial,ht3d-1kj8}. Most unusually, Phase 1, the ground state, is also multi-Q~\cite{wood2023double}. The magnetic structure is characterized by two distinct propagation vectors, in which a helix propagates along $\mathbf{q}_{1}$ and an orthogonal spin density wave (SDW) on $\mathbf{q}_{2}$ (see Fig.~\ref{fig:1} (b) for a schematic of incommensurate propagation vectors in \ce{GdRu2Si2}). Higher-order harmonics of the form $\mathbf{q}_1+2\mathbf{q}_2$ exist, on which a second magnetic helix propagates, compensating any varying spin density so that a constant moment structure forms. Preceding this study, scanning tunneling microscopy measurements of Yasui \textit{et al.}~\cite{yasui2020imaging}, being sensitive to the local density of states, revealed charge modulations at $2\mathbf{q}_{2}$ and $\mathbf{q}_{1}+\mathbf{q}_{2}$. Retrospectively, we note that the observation of $\mathbf{q}_{1}+\mathbf{q}_{2}$ satellites is decisive evidence that the two arms of the star are coupled together to form a multi-Q ground state. These CDWs arise as secondary order parameters (OPs) through magnetoelastic trilinear invariants in the free energy~\cite{izyumov1990phase}. The electronic structure has since been characterized by angle-resolved photoemission spectroscopy (ARPES)~\cite{eremeev2023insight,dong2025pseudogap} and quantum oscillations~\cite{matsuyama2023quantum}. Both ARPES studies identify Fermi surface nesting surrounding the X-point~\cite{hinuma2017band} matching the periodicity of $\mathbf{q}_{1}$, with a strongly peaked \ce{Gd} Lindhard function setting the Q-dependence of the RKKY interaction~\cite{dong2025pseudogap}. While Fermi surface nesting sets the magnetic periodicity, it cannot explain multi-Q ground state selection; neither density functional theory calculations~\cite{bouaziz2022fermi} nor inelastic neutron scattering~\cite{paddison2024spin,wood2025magnon} reveal higher-order or anisotropic exchange interactions.

In this Letter, we study the strong coupling between spin and charge degrees of freedom in \ce{GdRu2Si2}. In particular, we show how SDW components of the multi-Q ground state induce CDW order while helical components suppress it through destructive interference. Using polarized resonant elastic X-ray scattering (REXS) on a single rotational domain, we track the temperature dependence of the six magnetic and charge OPs associated with $\mathbf{q}_{1}$, $\mathbf{q}_{2}$, $2\mathbf{q}_{1}$, $2\mathbf{q}_{2}$, $\mathbf{q}_{1}+\mathbf{q}_{2}$ and $\mathbf{q}_{1}+2\mathbf{q}_{2}$. The interference is observed directly: cooling through $T_{1\leftrightarrow 4}$, the magnetic order on $\mathbf{q}_{1}$ becomes helical and the $2\mathbf{q}_{1}$ CDW is suppressed, whereas the $2\mathbf{q}_{2}$ CDW, driven by a collinear SDW on $\mathbf{q}_{2}$ continues to grow. A minimal Landau model, with one critical exponent $\beta$ and two ordering temperatures $T_{1\leftrightarrow 4}$ and $T_{\text{N}}$, reproduces the temperature dependencies of each OP. The coupling scheme constrains the high-temperature Phase 4 as a SDW phase, from which the ground state emerges by developing a helical component along one arm of the star. More generally, changes in the CDW structure between phases~\cite{yasui2020imaging,butler2026intra} not only introduce higher-order exchange into the effective Hamiltonian, but also suggest parametric changes in the Fermi-surface topography~\cite{matsuyama2023quantum,dong2025pseudogap}, mediating the interactions.

Samples were prepared as in Ref.~\cite{wood2023double}. Neutron experiments used $98.1\%$ isotopically enriched \ce{^{160}Gd} to mitigate absorption. Single-crystal and powder neutron diffraction experiments were performed on WISH at ISIS~\cite{chapon2011wish}. Polarized REXS measurements were performed on I16 at Diamond Light Source~\cite{collins2010diamond,abdeldaim2026one}. A single crystal was cleaved for a specular surface perpendicular to the $c$-axis. Incoming X-rays were tuned to the \ce{Gd} $L_{3}$ edge ($\SI{7.245}{\kilo\electronvolt}$) with $\sigma$-incident polarization and beam spot size of $\qtyproduct[product-units=repeat]{30 x 180}{\micro\meter}$. A \ce{Au} $(2,2,2)$ analyzer on a Stokes rotation stage resolved the outgoing polarization: $\sigma\rightarrow\sigma$ channel corresponds to charge scattering and $\sigma\rightarrow\pi$ to magnetic scattering~\cite{hill1996resonant}. Further experimental details are provided in the Supplemental  Material (SM)~\cite{SI}. 

We begin the analysis by examining the length of the primary magnetic propagation vectors in Phase 1 and Phase 4. \ce{GdRu2Si2} crystallizes in the tetragonal $I4/mmm$ space group, and the primary magnetic OPs transform as $m\Sigma$ type irreducible representations (irreps) which are associated with the $(q,0,0)$ propagation vector~\cite{aroyo2006bilbao1,aroyo2006bilbao2}.  Single crystal neutron diffraction measurements are shown in Fig.~\ref{fig:1} (c) at \SI{5}{\kelvin} in Phase 1 and \SI{43}{\kelvin} well within Phase 4. Phase 1 is characterized by two distinct propagation vectors $\mathbf{q}_1$ and $\mathbf{q}_2$~\cite{wood2023double}. With neutron diffraction, the two propagation vectors are observed along the same axis due to the two rotational domains (labeled with $A$ and $B$ subscripts) of the ground state. As temperature is raised these propagation vectors merge at $T_{1\leftrightarrow 4}$ so that $|\mathbf{q}_{1}| = |\mathbf{q}_{2}|$ in Phase 4. Therefore, in Phase 1, the magnetic OPs associated with $\mathbf{q}_{1}$ and $\mathbf{q}_{2}$ can have distinct temperature dependencies, whilst in Phase 4 they develop identically.

Next, REXS is used to identify the full set of incommensurate satellites in Phase 1 of \ce{GdRu2Si2}. Critically, a single rotational domain was probed in the ground state: scans along $h$ and $k$ each detect only one of $\mathbf{q}_{1}$ or $\mathbf{q}_{2}$ and only four $(\mathbf{q}_1 + 2\mathbf{q}_2)_{A}$ satellites are observed, with no intensity at the \SI{90}{\degree}-rotated $(\mathbf{q}_1 + 2\mathbf{q}_2)_{B}$ positions. Polarization analysis confirmed that odd-harmonic satellites ($\mathbf{q}_1$, $\mathbf{q}_2$ and $\mathbf{q}_1 + 2\mathbf{q}_2$) are exclusively magnetic ($\sigma\rightarrow\pi$), while even-harmonics ($2\mathbf{q}_1$, $2\mathbf{q}_2$ and  $\mathbf{q}_1 + \mathbf{q}_2$) are charge distortions ($\sigma\rightarrow\sigma$) [see SM~\cite{SI}]. These results are expected as higher-order harmonics of the primary magnetic distortions arise through trilinear coupling in the Landau free energy ($\mathcal{F}$). Preserving these invariants under time-reversal and translation symmetry requires that even harmonic terms are due to charge and odd harmonics to magnetism in the absence of a time-odd $q=0$ distortion. The trilinear coupling schemes available to the system are provided in SM~\cite{SI}.

\begin{figure}[h]
\includegraphics[width=0.8\linewidth,clip]{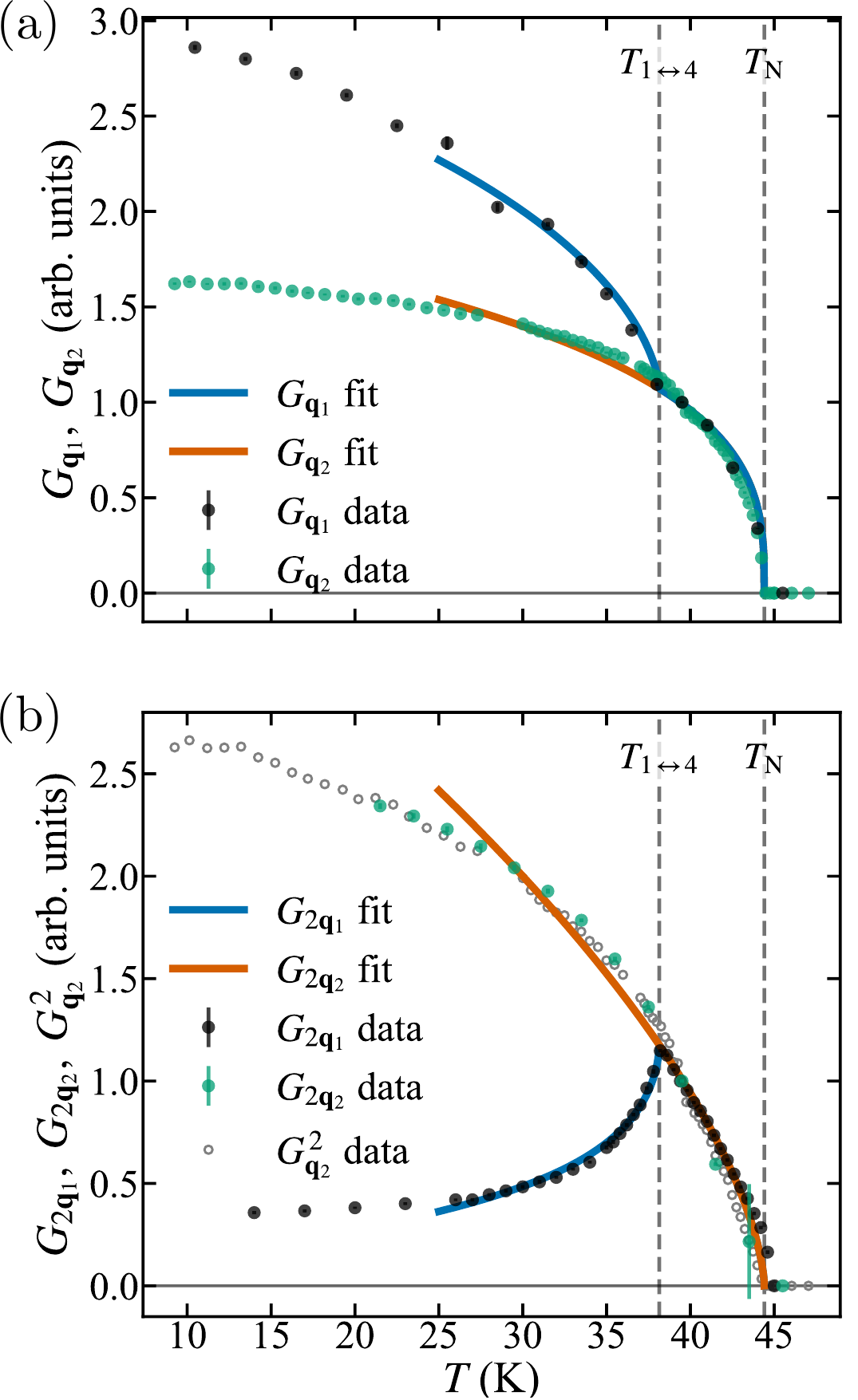}
\caption{\label{fig:2} Temperature dependencies of (a) $G_{\mathbf{q}_{1}}$ and $G_{\mathbf{q}_{2}}$ and (b) $G_{2\mathbf{q}_{1}}$, $G_{2\mathbf{q}_{2}}$ and $G_{\mathbf{q}_{2}}^2$ measured with REXS. Here $G_{\mathbf{q}}=\sqrt{I_{\mathbf{q}}}$ and $I_{\mathbf{q}}$ is the integrated intensity of a satellite at $\mathbf{q}$. Satellites were measured around the $(1,1,6)$ structural peak. A global fit of the Landau model described in the text has been applied to the data. }
\end{figure}

\begin{figure}[h]
\includegraphics[width=0.8\linewidth,clip]{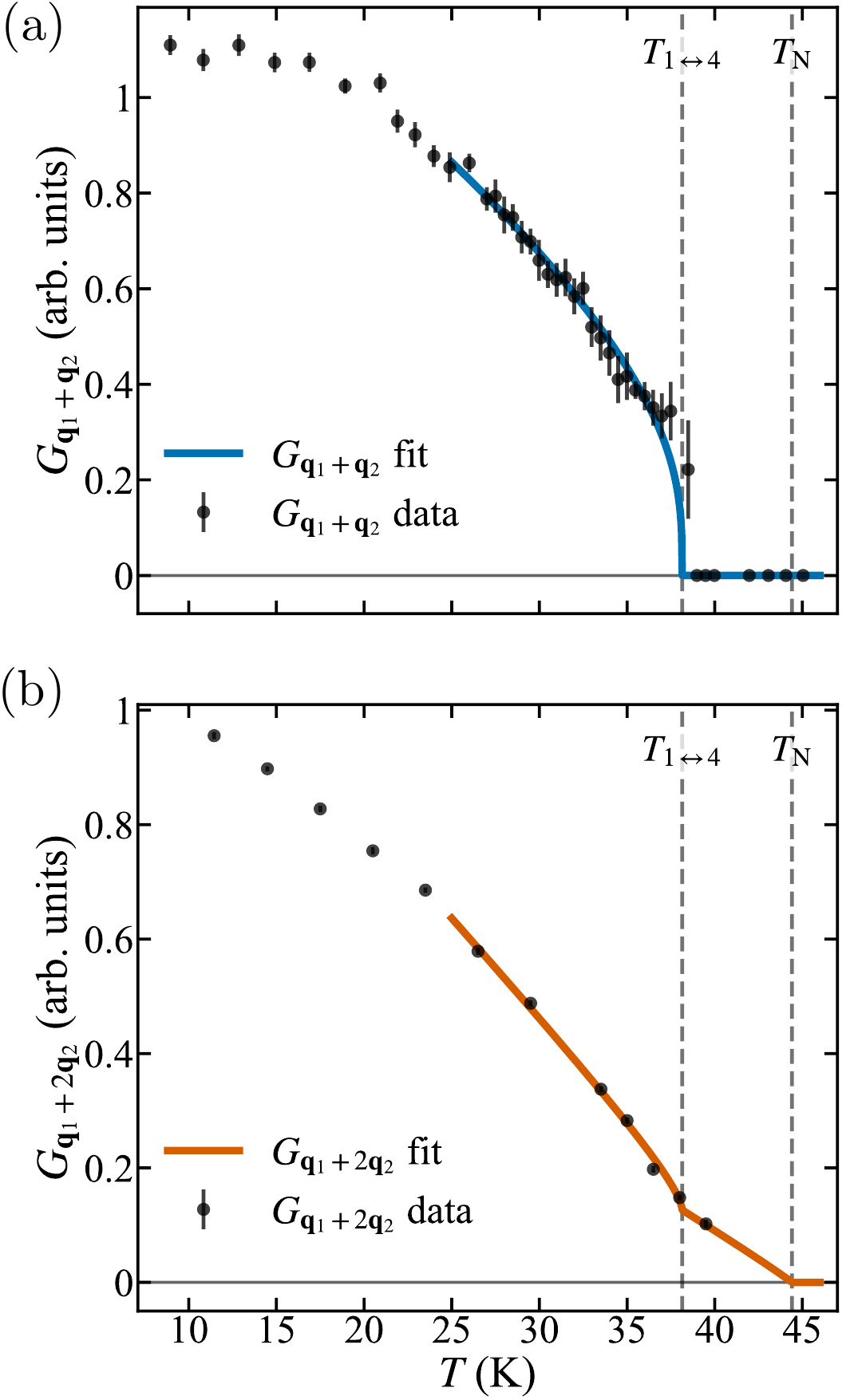}
\caption{\label{fig:3} Temperature dependencies of (a) $G_{\mathbf{q}_{1}+\mathbf{q}_{2}}$ and (b) $G_{\mathbf{q}_{1}+2\mathbf{q}_{2}}$  measured with REXS, where $G_{\mathbf{q}}=\sqrt{I_{\mathbf{q}}}$ and $I_{\mathbf{q}}$ is the integrated intensity of a satellite at $\mathbf{q}$. Satellites were measured around the $(1,1,6)$ structural peak. A global fit of the Landau model described in the text has been applied to the data.}
\end{figure}

Following this we examine the temperature dependencies of $G_{\mathbf{q}} = \sqrt{F_{\mathbf{q}} F^{*}_{\mathbf{q}}}$, with $F_{\mathbf{q}}$ denoting the structure factor at a given $\mathbf{q}$, for the incommensurate satellites. This quantity is closely related to the temperature dependencies of the various OPs. These are plotted in Figs.~\ref{fig:2} and~\ref{fig:3} as the square root of the intensities of incommensurate satellites surrounding the $(1, 1, 6)$ structural peak. To model these temperature dependencies, we use trilinear invariants to derive the temperature dependencies of non-primary OPs associated with $2\mathbf{q}_{1}$, $2\mathbf{q}_{2}$, $\mathbf{q}_{1}+\mathbf{q}_2$ and $\mathbf{q}_{1}+2\mathbf{q}_{2}$. The temperature dependence of an OP $[\rho(T)]$ introduced through a trilinear invariant can be derived by minimizing $\mathcal{F}$ with respect to $\rho$:

\begin{align}
\mathcal{F} &= \dots - \lambda_{1} \eta^2 \rho + \frac{\lambda_{2}}{2}\rho^2 + \dots \\
\rho(T) &= \frac{\lambda_{1}}{\lambda_{2}}\eta(T)^2,
\end{align}

\noindent where $\lambda_{1,2}$ are constants and $\eta(T)$ is an OP with known temperature dependence. We start by investigating the temperature dependencies of $G_{\mathbf{q}_{1}}(T)$, $G_{\mathbf{q}_{2}}(T)$ and the charge distortions $G_{2\mathbf{q}_{1}}$ and $G_{2\mathbf{q}_{2}}$ which are shown together in Fig.~\ref{fig:2}(a) and~\ref{fig:2}(b). These results are normalized at $T=\SI{39.5}{\kelvin}$ so that the temperature dependencies may be directly compared. Above $T_{1\leftrightarrow 4}$, in Phase 4, $G_{\mathbf{q}_{1}}(T)$ and $G_{\mathbf{q}_{2}}(T)$ map on to each other and likewise with $G_{2\mathbf{q}_{1}}(T)$ and $G_{2\mathbf{q}_{2}}(T)$. Below $T_{1\leftrightarrow 4}$, in Phase 1, the temperature dependencies diverge as permitted by the inequivalent propagation vectors. A simple temperature dependence for $G_{\mathbf{q}_2}(T)$ and $G_{2\mathbf{q}_{2}}(T)$ is first motivated. Both $G_{\mathbf{q}_2}(T)$ and $G_{2\mathbf{q}_{2}}(T)$ appear continuous between Phases 1 and 4. Given that $G_{\mathbf{q}_{2}}$ is due to a SDW in Phase 1~\cite{wood2023double}, this indicates that in Phase 4 both $G_{\mathbf{q}_{2}}$ and $G_{\mathbf{q}_{1}}$ correspond to SDWs. In developing models for the temperature dependencies, we adopt the notation,

 \begin{equation}
 \tau_{i} = \begin{cases}
 T_{i}-T & \text{if }T \leq T_{i}\\
 0 & \text{if } T > T_{i},\\
 \end{cases}
 \label{start}
 \end{equation}

  \noindent where the subscript $i = \{N,1\leftrightarrow 4\}$ denotes the two ordering temperatures. We posit that in the vicinity of $T_{\text{N}}$ the temperature dependence of $G_{\mathbf{q}_{2}}(T)$ is given by,
 
\begin{equation}
 G_{\mathbf{q}_{2}}(T)= A_{\mathbf{q}_{2}}\tau_{\text{N}}^{\beta}.
\end{equation}

Throughout, $A_{q}$ and $B_{q}$ refer to constants, with the subscripts denoting the associated propagation vector. Given that $G_{2\mathbf{q}_{2}}$ arises through a trilinear coupling to $G_{\mathbf{q}_{2}}$, its temperature dependence is entirely set through the primary magnetic distortion, such that,
   
\begin{equation}
G_{2\mathbf{q}_{2}}(T) =A_{2\mathbf{q}_2}\tau_{\text{N}}^{2\beta}.
\end{equation}

Indeed the temperature dependencies of $G_{\mathbf{q}_{2}}^2$ and $G_{2\mathbf{q}_{2}}$ map onto each other well [Fig.~\ref{fig:2}(b)], substantiating the coupling scheme. Subsequently, $G_{\mathbf{q}_{1}}$ may be considered; in Phase 4, this corresponds to the SDW of $G_{\mathbf{q}_{2}}$, but in Phase 1 it becomes the helical component of the ground state~\cite{wood2023double}. A magnetic helix is the superposition of two orthogonal SDWs with a relative phase difference of $\pi/2$; hence, the temperature dependence can be modeled as,

\begin{equation}
G_{\mathbf{q}_{1}}(T) =\sqrt{A^{2}_{\mathbf{q}_{1}}\tau_{\text{N}}^{2\beta} + B^{2}_{\mathbf{q}_{1}}\tau_{1\leftrightarrow 4}^{2\beta}}.
\label{eq_q1}
\end{equation}
 
$G_{\mathbf{q}_{1}}(T)$ may be interpreted as two SDWs with distinct ordering temperatures, which therefore grow differentially, which is essential to explain the temperature dependence of the charge order on $G_{2\mathbf{q}_{1}}(T)$. Below $T_{1\leftrightarrow 4}$, the development of $G_{2\mathbf{q}_{1}}$ is anticorrelated with $G_{2\mathbf{q}_{2}}$, showing a suppression in the charge ordering as temperature is lowered. To explain this, we consider that the three magnetic irreps ($m\Sigma_{2,3,4}$) responsible for $G_{\mathbf{q}_{1}}$ and $G_{\mathbf{q}_{2}}$ couple to the same CDW distortion ($\Sigma_{1}$) responsible for $G_{2\mathbf{q}_{1}}$ and $G_{2\mathbf{q}_{2}}$. Each SDW component forming the helix couples to CDW order at twice the magnetic wavevector, and therefore the two CDWs are $\pi$ out of phase and interfere destructively. The CDW on $G_{2\mathbf{q}_{1}}$ is not fully suppressed at the lowest measured temperatures. This is permitted as the coefficients for trilinear coupling involving the SDW components of the $b$-$c$ plane helix on $G_{\mathbf{q}_{1}}$ differ through the orthorhombic symmetry of the ground state. Therefore $G_{2\mathbf{q}_{1}}(T)$ is given by,

\begin{equation}
G_{2\mathbf{q}_{1}}(T) = \left| A_{2\mathbf{q}_{1}}\tau_{\text{N}}^{2\beta} - B_{2\mathbf{q}_{1}}\tau_{1\leftrightarrow 4}^{2\beta}\right|.
\end{equation}

We extend this approach to the multi-Q temperature dependencies of $G_{\mathbf{q}_{1}+\mathbf{q}_{2}}(T)$ and $G_{\mathbf{q}_{1}+2\mathbf{q}_{2}}(T)$ shown in Fig.~\ref{fig:3}(a) and~\ref{fig:3}(b). The onset of $G_{\mathbf{q}_{1}+\mathbf{q}_{2}}(T)$ is at $T_{1\leftrightarrow 4}$. This would suggest that mixed irreps ($m\Sigma_{3}$ and $m\Sigma_{4}$) between $G_{\mathbf{q}_{1}}$ and $G_{\mathbf{q}_{2}}$ couple together in Phase 1 to a $C_{2}$ irrep on $G_{\mathbf{q}_{1}+\mathbf{q}_{2}}$ so that,

\begin{equation}
G_{\mathbf{q}_{1}+\mathbf{q}_{2}}(T)=A_{\mathbf{q}_{1}+\mathbf{q}_{2}}\tau_{1\leftrightarrow 4}^{\beta}\tau_{\text{N}}^{\beta},
\end{equation}

 \noindent where the $\tau_{1\leftrightarrow 4}^\beta$ term is associated with the onset of one of the SDWs on $\mathbf{q}_{1}$ and the $\tau_{\text{N}}^\beta$ with $\mathbf{q}_{2}$. The $\mathbf{q}_{1}+2\mathbf{q}_{2}$ satellites are generated through a fourth-power invariant, however, equivalently it can be treated as a trilinear invariant of the form $\eta_{\mathbf{q}_{1}}\rho_{2\mathbf{q}_{2}} \eta_{\mathbf{q}_{1}+2\mathbf{q}_{2}}$ since the primary OPs couple to the $\Sigma_{1}$ irrep on $\rho_{2\mathbf{q}_{2}}$. Additionally, to generate the $b$-$c$ plane helix on $\eta_{\mathbf{q}_{1}+2\mathbf{q}_{2}}$ two trilinear coupling terms are required where $m\Sigma_{4}$ and $\Sigma_{1}$ couple to $mC_{1}$ to produce the $c$-axis SDW component of the helix and $m\Sigma_{3}$ and $\Sigma_{1}$ couple to $mC_{2}$ to form the $b$-axis SDW component. Therefore, the temperature dependence is given by,

 \begin{equation}
 G_{\mathbf{q}_{1}+2\mathbf{q}_{2}}(T) = \sqrt{ A^{2}_{\mathbf{q}_{1}+2\mathbf{q}_{2}}\tau_{1\leftrightarrow 4}^{2\beta}\tau_{\text{N}}^{4\beta}  + B^{2}_{\mathbf{q}_{1}+2\mathbf{q}_{2}}\tau_{\text{N}}^{6\beta} }.
 \label{Gq12q2}
 \end{equation}

Unlike the $\mathbf{q}_{1}+\mathbf{q}_{2}$ distortion, the $\mathbf{q}_{1}+2\mathbf{q}_{2}$ distortion remains non-zero above $T_{1\leftrightarrow 4}$, necessitating the $\tau^{6\beta}_{N}$ term in Eq.~\ref{Gq12q2}. Since $\mathbf{q}_{1}+2\mathbf{q}_{2}$ is a mixed harmonic, it can only arise where $\mathbf{q}_{1}$ and $\mathbf{q}_{2}$ coexist in the same volume, and not between single-Q domains. However, only one $\mathbf{q}_{1}+2\mathbf{q}_{2}$ peak could be measured in this region, owing to the weakness of the distortion near $T_{\text{N}}$. Tentatively, this result indicates that Phase 4 is double-Q. The magnetic structure of Phase 4 is studied in more detail in the End Matter~\ref{sec:endmatter} and comprises a tilted SDW model. This result is consistent with the small bump in resistivity observed in this phase, indicating a reduction in DOS at the Fermi surface~\cite{dong2025pseudogap}. The tilt of the SDW suggests competing sources of anisotropy, such as dipolar interactions~\cite{PhysRevB.103.064414,wood2025magnon,SarkarMech} and spin-orbit coupling to $5d$ electrons~\cite{colarieti2003origin,franse1980magnetic}. These may be important considerations in selecting a multi-Q state~\cite{LT_OG}.
 
 A fit of equations \ref{start}--\ref{Gq12q2} was applied to the data shown in Fig.~\ref{fig:2} and Fig.~\ref{fig:3}, with the three global parameters converging on $\beta=\num{0.319\pm0.008}$, $T_N=\qty{44.41\pm0.07}{K}$ and $T_{1\leftrightarrow 4}=\qty{38.14\pm0.09}{K}$. We note that the two ordering temperatures are systematically lower ($\sim \qty{1.5}{\kelvin}$) than determinations by thermodynamic measurements~\cite{gries2025uniaxial}; this is explained by beam heating effects expected at the surface of the sample. Furthermore, $\beta$ acts as a shape parameter for the model. The fit, applied for $T>\SI{25}{\kelvin}$ extends well beyond the asymptotic critical region of either transition at $T_{1\leftrightarrow 4}$ and $T_{\text{N}}$, and therefore $\beta$ should not be used to assign a universality class~\cite{pelissetto2002critical}.

In summary, our study provides a phenomenological coupling scheme for multi-Q magnetic and charge OPs in \ce{GdRu2Si2}. We show through polarized REXS that SDWs promote CDW order, and that magnetic helices suppress it through destructive interference. The evolution of the OPs between Phases 1 and 4 shows that Phase 4 is a tilted SDW phase, which is indicative of intrinsic sources of anisotropy. Moreover, concomitant magnetic and CDW order introduces higher-order interactions to the system (such as biquadratic exchange) via a renormalized Hamiltonian~\cite{PhysRev.120.335}. This is significant as, while bilinear Heisenberg exchange is the dominant magnetic energy scale~\cite{paddison2024spin,wood2025magnon}, ultimately it is anisotropic and higher-order exchange terms that are decisive in selecting multi-Q ground states in centrosymmetric systems. Additionally, temperature- and field-dependent changes in the CDW structure~\cite{yasui2020imaging,PhysRevB.104.144404,butler2026intra} feed back into the Fermi-surface topography and effective spin Hamiltonian. Microscopic models of multi-$Q$ intermetallics must therefore explicitly account for the evolution and symmetry of coupled CDW order; our analysis of \ce{GdRu2Si2} provides a rigorous benchmark in that regard.

\begin{acknowledgments}
We would like to acknowledge useful discussions with Julie Staunton and Paul Goddard (University of Warwick) and Joseph Paddison (ORNL). This work was financially supported by two Engineering and Physical Sciences Research Council grants: EP/T005963/1, and the UK Skyrmion Project Grant, EP/N032128/1. We acknowledge Diamond Light Source for time on beamline I16 under proposal MM30726, and ISIS Neutron and Muon Source for time on the WISH instrument under proposals RB2210259 and RB2310320 and partial funding of the \ce{^160Gd} isotope~\cite{Gdata_WISH1,Gdata_WISH2}.
\end{acknowledgments}

\bibliographystyle{apsrev4-2}
\bibliography{refs}

\section{End Matter}
\label{sec:endmatter}

\begin{figure}[b]
\includegraphics[width=1\linewidth,clip]{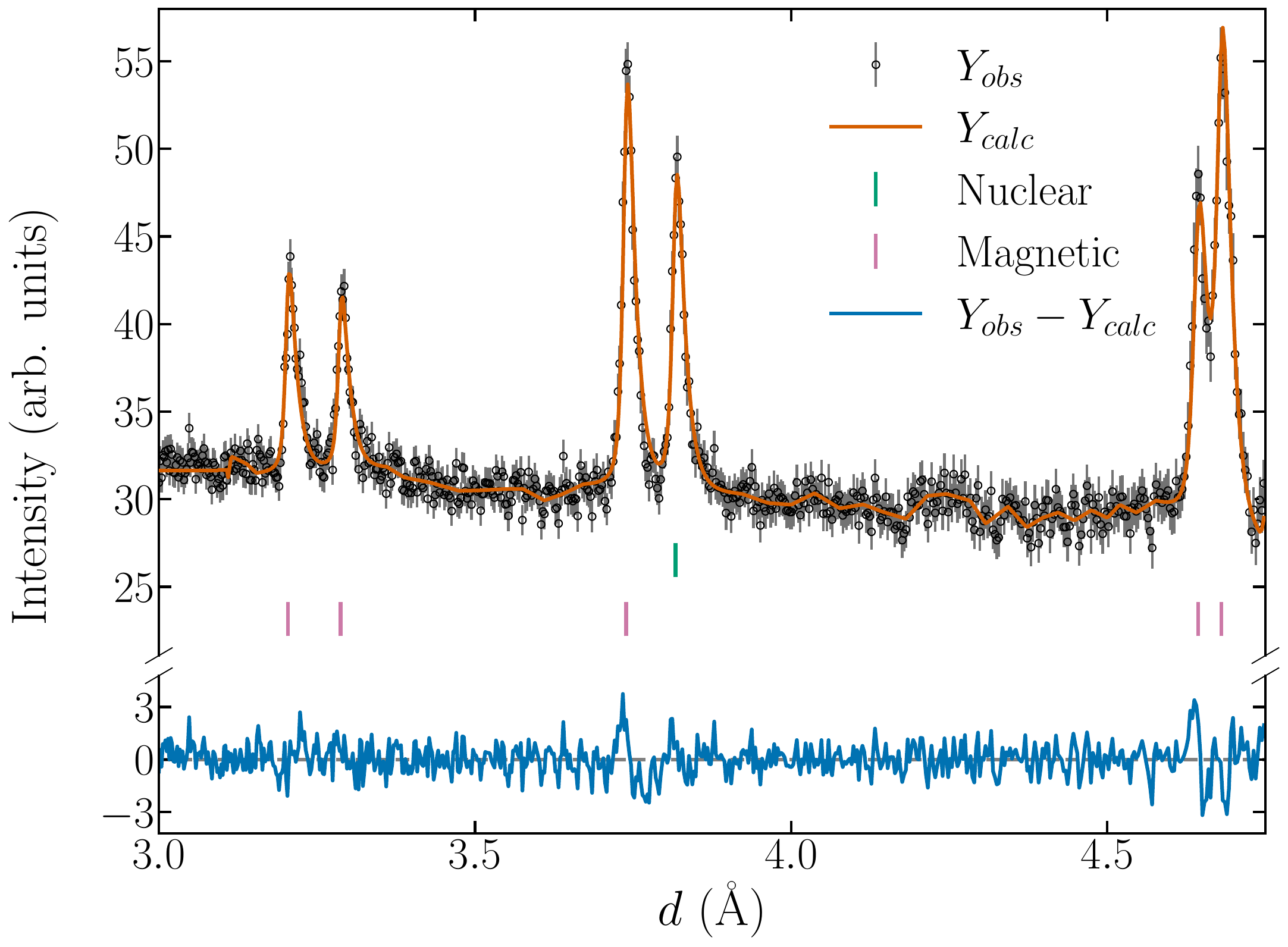}
\caption{\label{fig:EM} Rietveld refinement using a tilted SDW model of the powder neutron diffraction data for \ce{GdRu2Si2} collected in Phase 4 at $T=\SI{40}{\kelvin}$. The moment direction is in the ($\phi=\SI{66.2}{\degree},\theta=\SI{45.9}{\degree}$) direction where $\phi$ is the angle of the moment in the $a$-$b$ plane measured from the $a$-axis and $\theta$ the angle between the moment and the $c$-axis. The magnetic structure is formed from the admixture of three irreps $m\Sigma_{2}$, $m\Sigma_{3}$ and $m\Sigma_{4}$. Higher-order harmonics are not evident in the powder neutron diffraction due to the strong absorption~\cite{wood2023double}. \textsc{fullprof} fitting statistics: $R_{wp}=22.6\%$, $R_{exp}=25.9\%$ and $\chi^2=0.76$~\cite{rodriguez1993recent}.}
\end{figure}

As asserted previously, the coupling of magnetic order to the CDWs in Phase 4 implies SDW order. Fitting SDW models to powder neutron diffraction data taken in Phase 4 yields the optimal solution shown in Fig.~\ref{fig:EM}, which consists of an admixture of three irreps, $m\Sigma_{2}$, $m\Sigma_{3}$, and $m\Sigma_{4}$, placing the moment axis of the SDW in a general direction. Note that in this case, powder neutron diffraction cannot distinguish between the single- and double-Q SDW models. Single- and two-irrep solutions result in inferior fits and are shown in the SM~\cite{SI}. This is consistent with the refinements of Paddison \textit{et al.,} which identified the three-irrep solution as the highest quality~\cite{paddison2024spin}. A three-irrep solution is unusual: second-order phase transitions from the paramagnetic phase ordinarily decompose into a single-irrep. However, recent dilatometric and heat capacity measurements revealed Phase 6 occupying a narrow ($\sim\qty{1}{\kelvin}$) region between Phase 4 and the paramagnetic state~\cite{gries2025uniaxial}. No signatures of Phase 6 are resolved in the REXS temperature dependence, its narrow width being comparable to our temperature-step resolution near $T_{\text{N}}$. A magnetic structure determination of Phase 6 is also precluded by insufficient counting statistics in proximity to $T_{\text{N}}$. We posit that \ce{GdRu2Si2} decomposes into a single irrep in Phase 6, with additional irreps emerging through cooling into Phase 4. Powder neutron diffraction cannot determine the relative phases of these three irreps, but the coupling to CDW order in Phase 4 strongly indicates that they develop in phase. A general moment direction is not unusual in metallic \ce{Gd} systems, where dipolar interactions compete with $5d$ spin-orbit anisotropy~\cite{colarieti2003origin,franse1980magnetic}. Since the constant-moment ansatz of the ground state is basis independent~\cite{wood2023double}, the tilted SDW of Phase 4 could suggest that the ground state itself adopts a low-symmetry basis. A search over 500 randomly oriented orthonormal bases (see SM~\cite{SI}) discounts this possibility, recovering the $b$-$c$ plane helix on $\mathbf{q}_{1}$ with an orthogonal SDW on $\mathbf{q}_{2}$~\cite{wood2023double,paddison2024spin}.

\end{document}


\title{Destructive Interference of a Charge Density Wave in the Skyrmion Magnet \ce{GdRu2Si2}}

\author{G. D. A. Wood}
\email{george.wood@stfc.ac.uk}
\affiliation{ISIS Neutron and Muon Source, STFC Rutherford Appleton Laboratory, Harwell Science and Innovation Campus, United Kingdom}
\affiliation{Department of Physics, University of Warwick, Coventry, CV4 7AL, United Kingdom}

\author{D. A. Mayoh}
\affiliation{Department of Physics, University of Warwick, Coventry, CV4 7AL, United Kingdom}

\author{F. Orlandi}
\affiliation{ISIS Neutron and Muon Source, STFC Rutherford Appleton Laboratory, Harwell Science and Innovation Campus, United Kingdom}

\author{A. Bombardi}
\affiliation{Diamond Light Source Ltd, Harwell Science and Innovation Campus, Didcot, Oxfordshire, United Kingdom}

\author{D. D. Khalyavin}
\affiliation{ISIS Neutron and Muon Source, STFC Rutherford Appleton Laboratory, Harwell Science and Innovation Campus, United Kingdom}

\author{P. Manuel}
\affiliation{ISIS Neutron and Muon Source, STFC Rutherford Appleton Laboratory, Harwell Science and Innovation Campus, United Kingdom}

\author{S. H. Moody}
\affiliation{Laboratory for Neutron Scattering and Imaging (LNS), PSI Center for Neutron and Muon Science, Paul Scherrer Institut (PSI), CH-5232 Villigen, Switzerland}
\affiliation{Department of Physics, Durham University, South Road, Durham DH1 3LE, United Kingdom}

\author{P. D. Hatton}
\affiliation{Department of Physics, Durham University, South Road, Durham DH1 3LE, United Kingdom}

\author{R. D. Johnson}
\affiliation{Department of Physics, Durham University, South Road, Durham DH1 3LE, United Kingdom}

\author{T. W. Chamberlain}
\affiliation{Department of Physics, University of Warwick, Coventry, CV4 7AL, United Kingdom}

\author{O. A. Petrenko}
\affiliation{Department of Physics, University of Warwick, Coventry, CV4 7AL, United Kingdom}

\author{M. R. Lees}
\affiliation{Department of Physics, University of Warwick, Coventry, CV4 7AL, United Kingdom}

\author{G. Balakrishnan}
\email{g.balakrishnan@warwick.ac.uk}
\affiliation{Department of Physics, University of Warwick, Coventry, CV4 7AL, United Kingdom}
\date{\today}
\maketitle


%
%
\section{Additional Experimental Details}

For single-crystal neutron diffraction the sample was mounted on an aluminum
strip holder with $a^{*}$--$b^{*}$ as the principal scattering plane. For the powder
measurements, polycrystalline material was finely ground and loaded into a
vanadium can. Both experiments were performed in a helium cryostat with a base
temperature of \SI{1.5}{\kelvin}. For the REXS experiment, the cleaved single
crystal was affixed to a copper holder, encased in a beryllium dome and cooled
with a closed-cycle refrigerator to a base temperature of \SI{7}{\kelvin}. The
diffractometer was operated in a vertical scattering geometry, so that the
horizontally polarized incident beam is $\sigma$ polarized with respect to the
scattering plane. Outgoing polarization was resolved with a single-crystal gold
$(2,2,2)$ analyzer mounted on a Stokes rotation stage.

The temperature dependencies of satellite intensities surrounding the $(1,1,6)$ structural peak were measured using polarized REXS. The following satellites were measured: $(1+q_{1},1,6)$, $(1,1+q_{2},6)$, $(1+2q_{1},1,6)$, $(1,1+2q_{2},6)$, $(1-q_{1},1+q_{2},6)$ and $(1+q_{1},1+2q_{2},6)$ where $q_{1}=0.2200$ and $q_{2}=0.2246$. Note that the notation $\mathbf{q}_{1}+\mathbf{q}_{2}$ throughout the text denotes the mixed harmonic satellite generically; the four satellites of this set $(1\pm q_{1},1\pm q_{2},6)$ share the same temperature dependence by symmetry, but the intensities differ due to structure-factor differences. The $\mathbf{q}_{1}+\mathbf{q}_{2}$ peaks were exceptionally weak, and of these the $(1-q_{1},1+q_{2},6)$ satellite was the strongest of this type. Therefore, we measured the temperature dependence of this satellite, but quote $G_{\mathbf{q}_{1}+\mathbf{q}_{2}}$ as being representative throughout. The temperature dependencies of the $\mathbf{q}_{1}$, $\mathbf{q}_{2}$ and $\mathbf{q}_{1}+2\mathbf{q}_{2}$ satellites were measured through the $\sigma\rightarrow\pi$ (magnetic scattering) channel. The $2\mathbf{q}_{1}$ and $2\mathbf{q}_{2}$ satellites were measured in the $\sigma\rightarrow\sigma$ (charge scattering) channel. The $\mathbf{q}_{1}+\mathbf{q}_{2}$ satellite temperature dependence could not be consistently measured through the $\sigma\rightarrow\sigma$ channel due to the weakness of the signal in proximity to $T_{14}$. Hence, the temperature dependence for this satellite was measured without outgoing polarization analysis to minimize flux loss. This resulted in a noisier measurement; therefore, the error bars in $G_{\mathbf{q}_{1}+\mathbf{q}_{2}}$ are larger than the other temperature dependencies in Fig. 3 (a) of the main text.

\newpage

\section{Polarized REXS Results}

In this section, we determine whether a given peak is due to charge or magnetic scattering using polarized REXS as described in the main text. We performed $\eta$ rocking scans (sample rotation parallel to the beam~\cite{abdeldaim2026one}) over a given peak whilst measuring intensity through a given polarization channel ($\sigma\rightarrow\pi$ or $\sigma\rightarrow\sigma$). The results confirm that the structural $(1,1,6)$ peak is due to charge scattering, the  $(1+q_{1},1,6)$, $(1,1+q_{2},6)$ and $(1+q_{1},1+2q_{2},6)$ peaks (odd harmonics) are due to magnetic scattering and that the $(1+2q_{1},1,6)$, $(1,1+2q_{2},6)$ and $(1+q_{1},1+q_{2},6)$ peaks (even harmonics) are due to charge scattering~\cite{hill1996resonant}. The peaks were fit with a pseudo-Voigt function. The peaks have a consistent full width at half maximum, with $\text{FWHM } = (\num{8.87(7)e-3})\si{\degree}$. This substantiates that the correlation lengths of the magnetic and charge orders arise as part of a unified ground state, not as distinct processes.

\begin{figure}[h]
\centering
\includegraphics[width=1\linewidth]{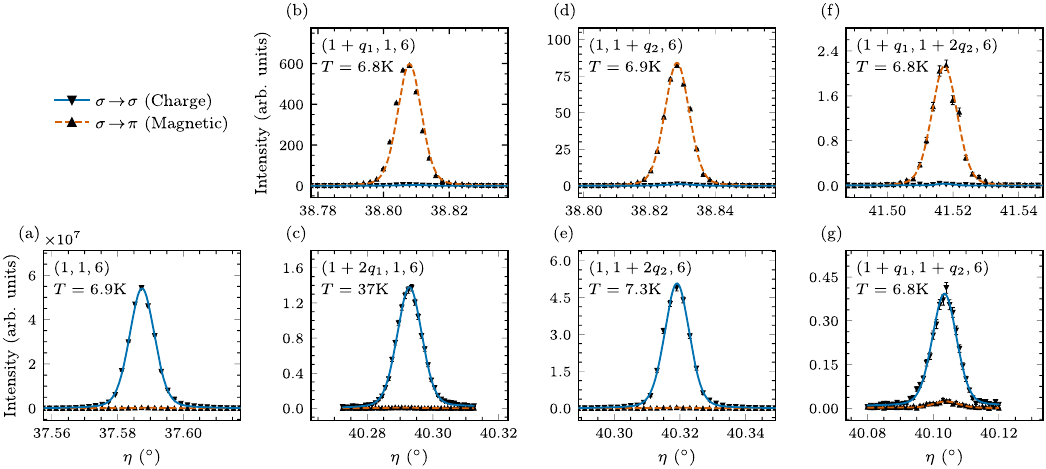}
\caption{Polarized REXS results: $\eta$ rocking angle versus intensity of (a) $(1,1,6)$ (b) $(1+q_{1},1,6)$ (c) $(1+2q_{1},1,6)$  (d) $(1,1+q_{2},6)$, (e) $(1,1+2q_{2},6)$, (f) $(1+q_{1},1+2q_{2},6)$ and (g) $(1+q_{1},1+q_{2},6)$. These measurements took place at base temperature ($\sim\num{7}\si{\kelvin}$), apart from the $(1+2q_{1},1,6)$ peak, which was measured at $T=\qty{37}{\kelvin}$, as at this temperature the CDW is strongest, before being suppressed at lower temperatures.}
\label{fig:S1}
\end{figure}

\newpage
\section{Trilinear Couplings in \NoCaseChange{\ce{GdRu2Si2}}}

We consider the permitted trilinear couplings that form between order parameters transforming as $(m)\Sigma$ and $(m)C$ type irreps~\cite{aroyo2006bilbao1,aroyo2006bilbao2}. The vectors of the star for the $m\Sigma$ and $\Sigma$ irreps are: $\mathbf{k}_{1}=(u,0,0)$, $\mathbf{k}_{2}=(-u,0,0)$, $\mathbf{k}_{3}=(0,-u,0)$ and $\mathbf{k}_{4}=(0,u,0)$. The vectors of the star for the $mC$ and $C$ irreps are: $\mathbf{k}_{1}=(u,v,0)$, $\mathbf{k}_{2}=(-u,v,0)$, $\mathbf{k}_{3}=(v,-u,0)$, $\mathbf{k}_{4}=(-v,u,0)$, $\mathbf{k}_{5}=(-u,-v,0)$, $\mathbf{k}_{6}=(u,-v,0)$, $\mathbf{k}_{7}=(v,u,0)$, $\mathbf{k}_{8}=(-v,-u,0)$.

Trilinear invariants, bilinear in two time-odd irreps and linear in one time-even irrep, must be invariant under the generators of the space group ($I4/mmm$) to be valid: $\{1|t_{1},t_{2},t_{3}\}$, $\{2_{001}|0\}$, $\{4^{+}_{001}|0\}$, $\{2_{010}|0\}$, $\{-1|0\}$ and $\{1'|0\}$. For brevity, we list the permitted and forbidden irrep combinations rather than the full invariant expressions in terms of order parameter components.

\begin{table}[h]
\centering
\caption{Trilinear invariants permitted and forbidden by the symmetry of
\NoCaseChange{\ce{GdRu2Si2}} (space group $I4/mmm$).}
\label{tab:invariants}
\begin{tabular}{@{}c@{\hspace{2em}}|@{\hspace{2em}}c@{}}
\toprule
Permitted & Forbidden \\
\midrule
\multicolumn{2}{c}{$m_{\mathbf{q}_1}$ coupling to $\rho_{\mathbf{q}_1}$ and} \\
\multicolumn{2}{c}{$m_{\mathbf{q}_2}$ coupling to $\rho_{\mathbf{q}_2}$} \\
\midrule
$(m\Sigma_{2})^2 \Sigma_1$ & $(m\Sigma_{2})^2 \Sigma_2$ \\
$(m\Sigma_{3})^2 \Sigma_1$ & $(m\Sigma_{3})^2 \Sigma_2$ \\
$(m\Sigma_{4})^2 \Sigma_1$ & $(m\Sigma_{4})^2 \Sigma_2$ \\
 & $(m\Sigma_{2})^2 \Sigma_3$ \\
 & $(m\Sigma_{3})^2 \Sigma_3$ \\
 & $(m\Sigma_{4})^2 \Sigma_3$ \\
 & $(m\Sigma_{2})^2 \Sigma_4$ \\
 & $(m\Sigma_{3})^2 \Sigma_4$ \\
 & $(m\Sigma_{4})^2 \Sigma_4$ \\
$(m\Sigma_2)(m\Sigma_3)\Sigma_4$  & $(m\Sigma_2)(m\Sigma_3)\Sigma_1$ \\
 & $(m\Sigma_2)(m\Sigma_3)\Sigma_2$ \\
 & $(m\Sigma_2)(m\Sigma_3)\Sigma_3$ \\
$(m\Sigma_2)(m\Sigma_4)\Sigma_3$ & $(m\Sigma_2)(m\Sigma_4)\Sigma_1$ \\
 & $(m\Sigma_2)(m\Sigma_4)\Sigma_2$ \\
 & $(m\Sigma_2)(m\Sigma_4)\Sigma_4$ \\
$(m\Sigma_3)(m\Sigma_4)\Sigma_2$ & $(m\Sigma_3)(m\Sigma_4)\Sigma_1$ \\
 & $(m\Sigma_3)(m\Sigma_4)\Sigma_3$ \\
 & $(m\Sigma_3)(m\Sigma_4)\Sigma_4$ \\
\midrule
\multicolumn{2}{c}{Coupling between $m_{\mathbf{q}_1}$,$m_{\mathbf{q}_{2}}$ and $\rho_{\mathbf{q}_{1} + \mathbf{q}_{2}}$} \\
\midrule
$(m\Sigma_{2})^2 C_{1}$ & $(m\Sigma_{2})^2 C_{2}$ \\
$(m\Sigma_{3})^2 C_{1}$ & $(m\Sigma_{3})^2 C_{2}$ \\
$(m\Sigma_{4})^2 C_{1}$ & $(m\Sigma_{4})^2 C_{2}$ \\
$m\Sigma_{2} m\Sigma_{3} C_{1}$ & $m\Sigma_{2} m\Sigma_{3} C_{2}$ \\
$m\Sigma_{2} m\Sigma_{4} C_{2}$ & $m\Sigma_{2} m\Sigma_{4} C_{1}$ \\
$m\Sigma_{3} m\Sigma_{4} C_{2}$ & $m\Sigma_{3} m\Sigma_{4} C_{1}$ \\ 
\midrule
\multicolumn{2}{c}{Coupling between $m_{\mathbf{q}_{1}}$, $\rho_{2\mathbf{q}_{2}}$ and $m_{\mathbf{q}_{1}+2\mathbf{q}_{2}}$} \\
\midrule
$m\Sigma_{4} \Sigma_{1} mC_{1}$ & $m\Sigma_{4} \Sigma_{1} mC_{2}$ \\
$m\Sigma_{2} \Sigma_{1} mC_{2}$ & $m\Sigma_{2} \Sigma_{1} mC_{1}$ \\
$m\Sigma_{3} \Sigma_{1} mC_{2}$ & $m\Sigma_{3} \Sigma_{1} mC_{1}$ \\
\bottomrule
\end{tabular}
\end{table}

\newpage
\clearpage

\section{Powder Neutron Diffraction Models of Phase 4}

\begin{figure*}[h]
  \centering
  \captionsetup[subfigure]{skip=4pt, margin=0pt}
  \setlength{\abovecaptionskip}{4pt}
  \setlength{\belowcaptionskip}{0pt}

  \begin{subfigure}[t]{0.32\textwidth}
    \includegraphics[width=\linewidth]{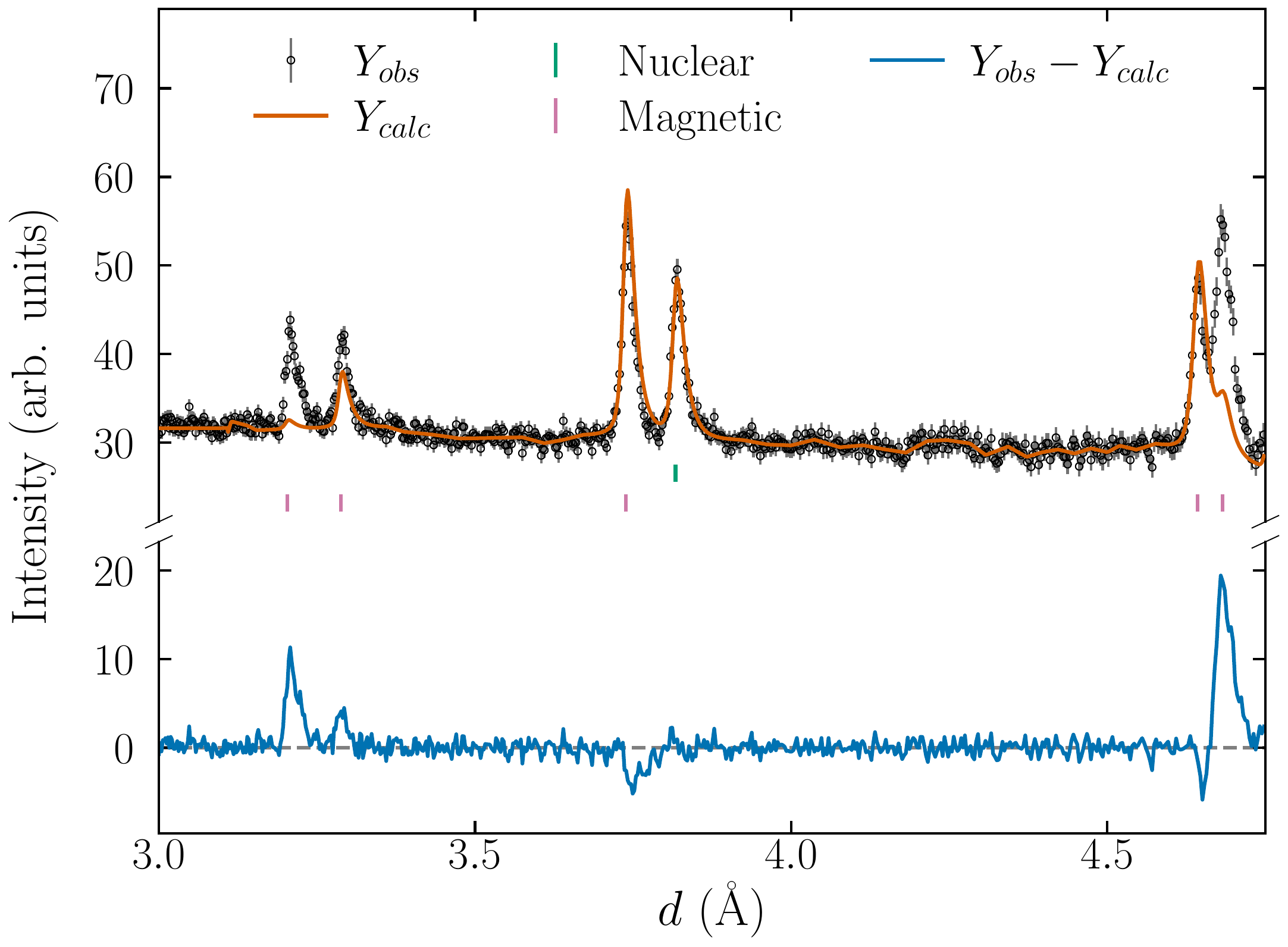}
    \caption{SDW model with moments along the $a$-axis ($\phi=\SI{0}{\degree}$, $\theta=\SI{90}{\degree}$). Magnetic structure is formed from the $m\Sigma_{2}$ irrep. \textsc{fullprof} fitting statistics: $R_{wp}=51.0\%$ and $\chi^{2}=3.859$}\label{fig:si-a}
  \end{subfigure}\hfill
  \begin{subfigure}[t]{0.32\textwidth}
    \includegraphics[width=\linewidth]{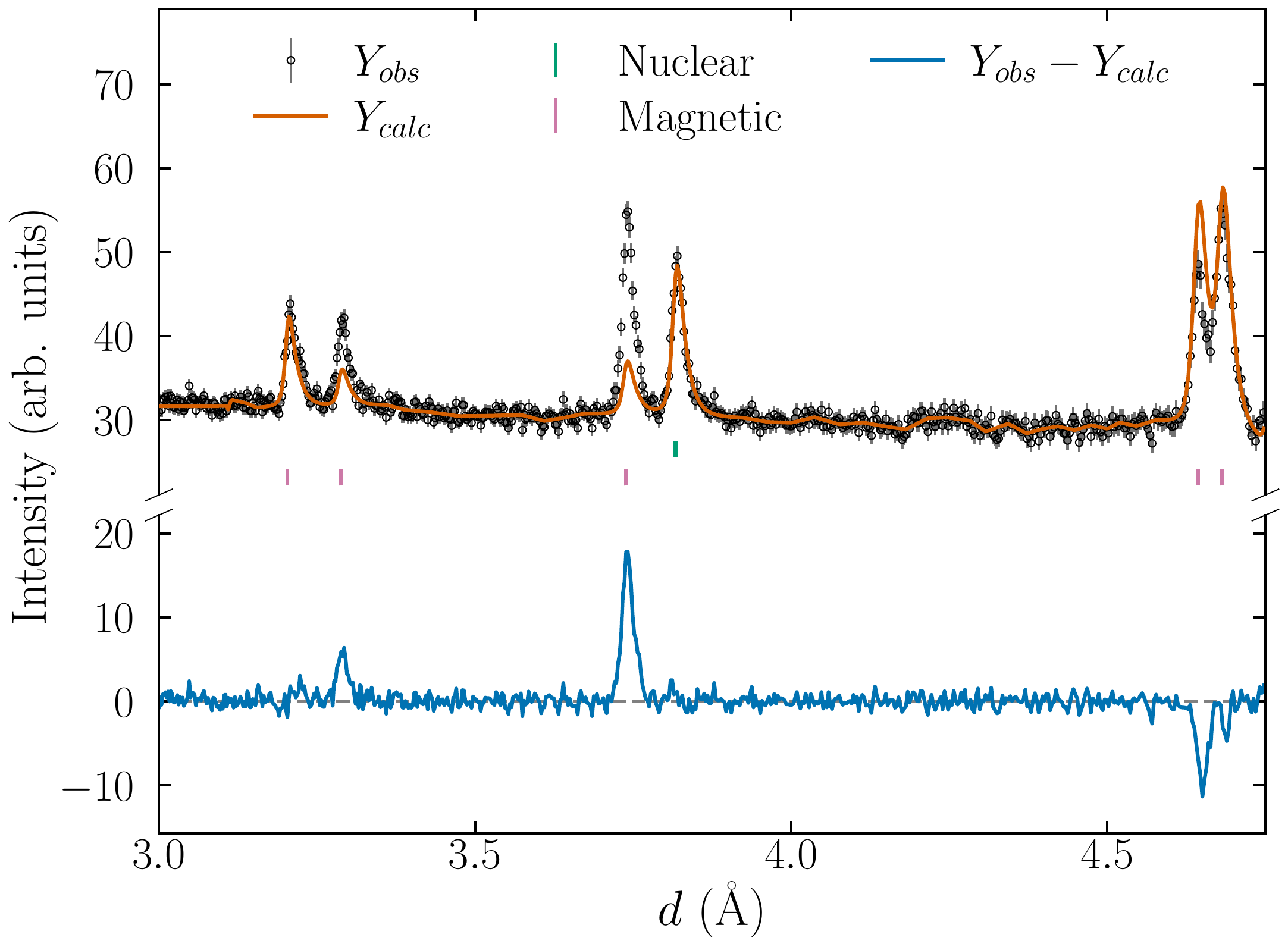}
    \caption{SDW model with moments along the $b$-axis ($\phi=\SI{90}{\degree}$, $\theta=\SI{90}{\degree}$). Magnetic structure is formed from the $m\Sigma_{3}$ irrep. \textsc{fullprof} fitting statistics: $R_{wp}=48.8\%$ and $\chi^{2}=3.539$}\label{fig:si-b}
  \end{subfigure}\hfill
  \begin{subfigure}[t]{0.32\textwidth}
    \includegraphics[width=\linewidth]{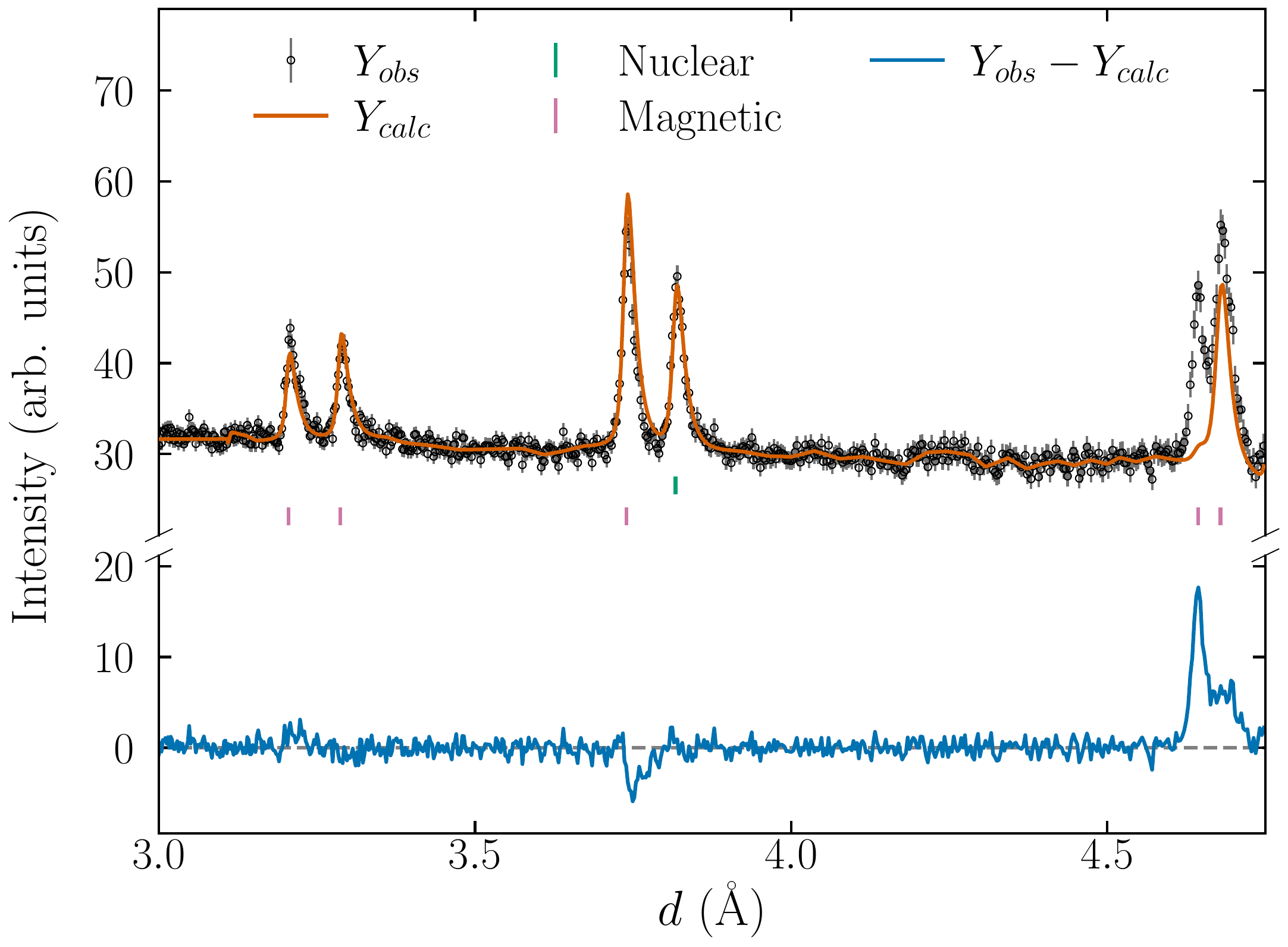}
    \caption{SDW model with moments along the $c$-axis ($\theta=\SI{0}{\degree}$). Magnetic structure is formed from the $m\Sigma_{4}$ irrep. \textsc{fullprof} fitting statistics: $R_{wp}=39.8\%$ and $\chi^{2}=2.358$}\label{fig:si-c}
  \end{subfigure}
  
  \vspace{1em} 

  \begin{subfigure}[t]{0.32\textwidth}
    \includegraphics[width=\linewidth]{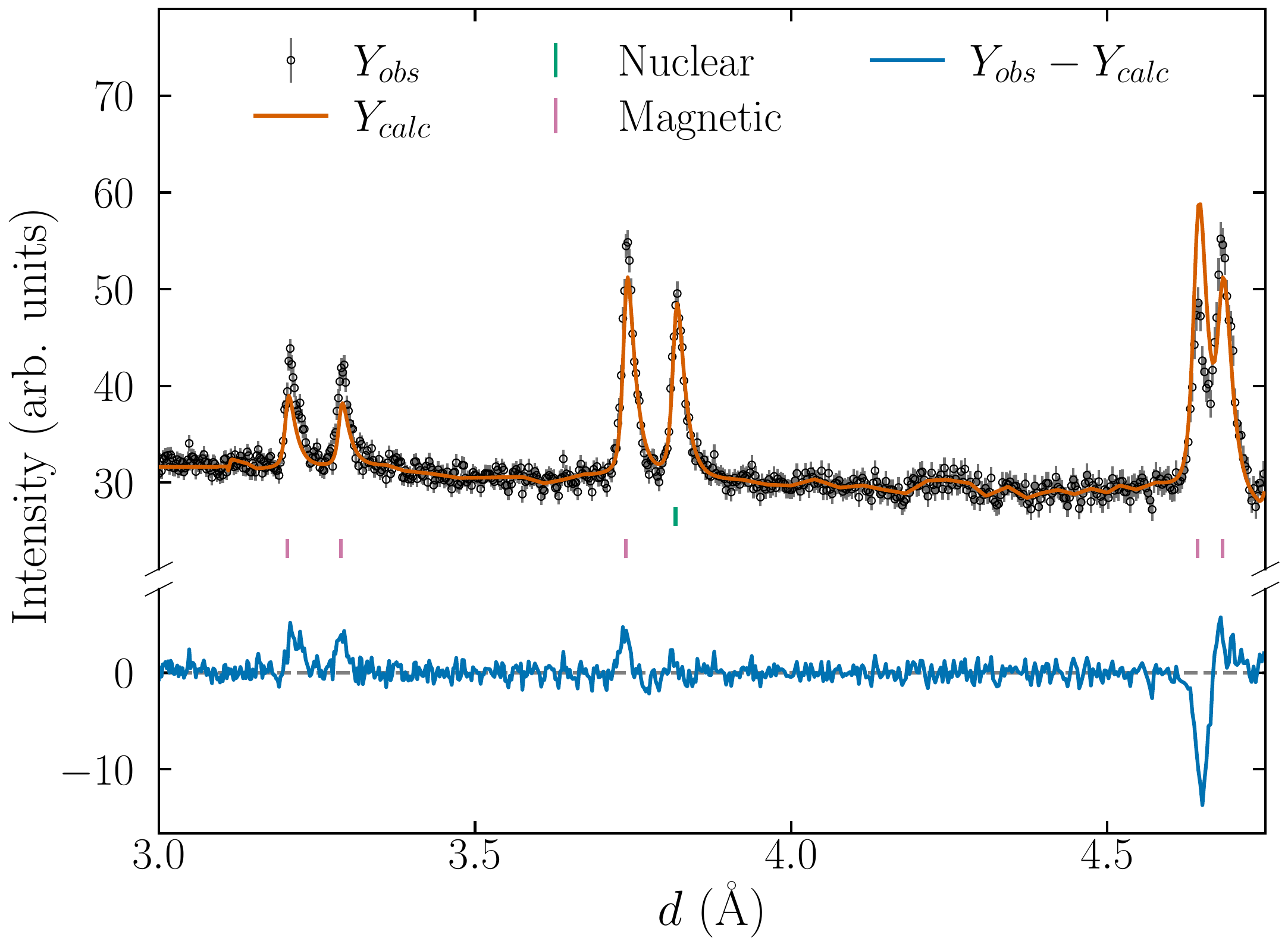}
    \caption{SDW model with moments along ($\phi=\SI{47.8}{\degree}$, $\theta=\SI{90}{\degree}$), \textit{i.e.} within $a$-$b$ plane. Magnetic structure is formed from an admixture of two irreps: $m\Sigma_{2}$ and $m\Sigma_{3}$. \textsc{fullprof} fitting statistics: $R_{wp}=34.3\%$ and $\chi^{2}=1.748$}\label{fig:si-d}
  \end{subfigure}\hfill
  \begin{subfigure}[t]{0.32\textwidth}
    \includegraphics[width=\linewidth]{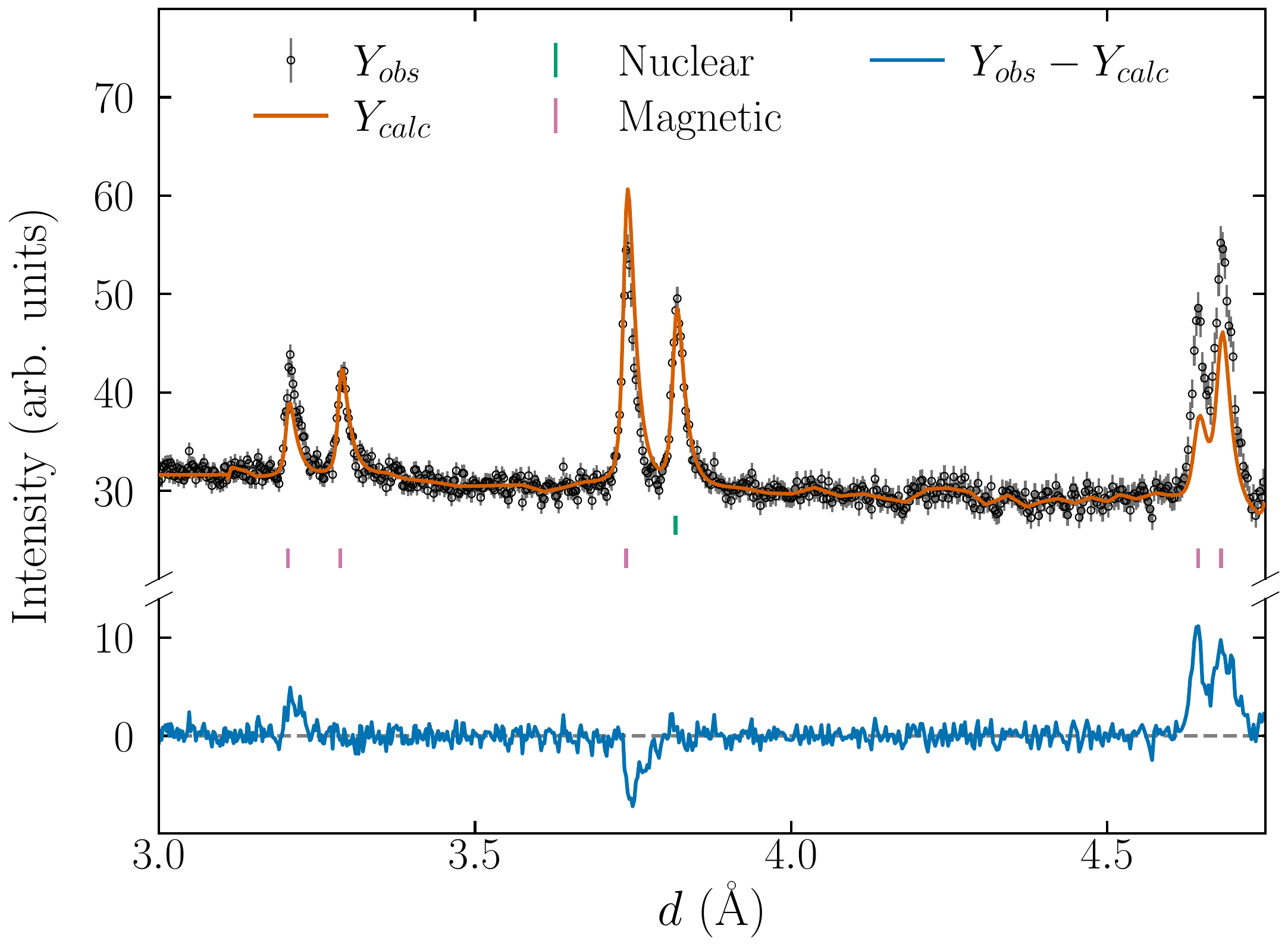}
    \caption{SDW model with moments along ($\phi=\SI{0}{\degree}$, $\theta=\SI{31.7}{\degree}$), \textit{i.e.} within $a$-$c$ plane. Magnetic structure is formed from an admixture of two irreps: $m\Sigma_{2}$ and $m\Sigma_{4}$. \textsc{fullprof} fitting statistics: $R_{wp}=37.8\%$ and $\chi^{2}=2.125$}\label{fig:si-e}
  \end{subfigure}\hfill
  \begin{subfigure}[t]{0.32\textwidth}
    \includegraphics[width=\linewidth]{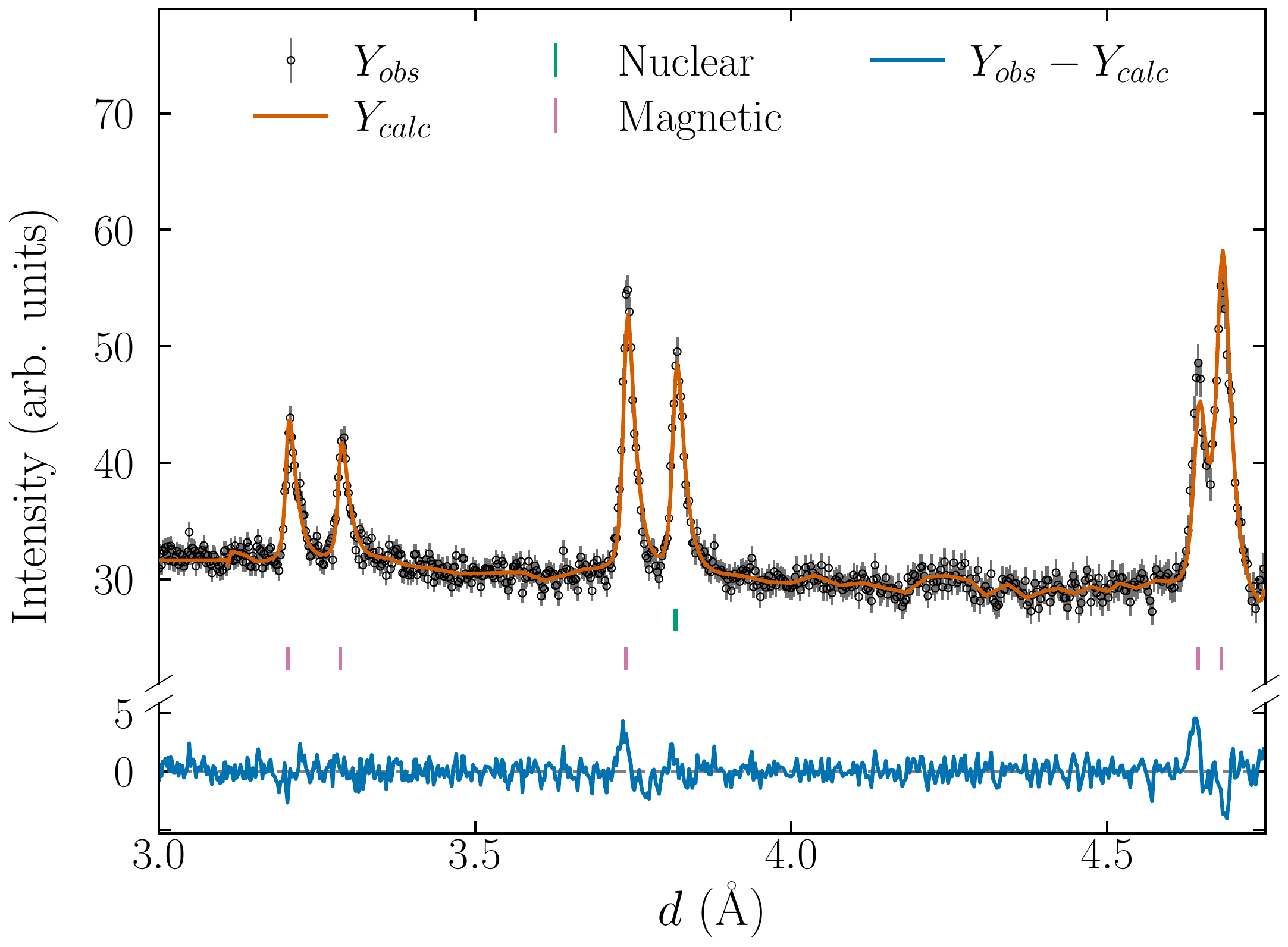}
    \caption{SDW model with moments along ($\phi=\SI{90}{\degree}$, $\theta=\SI{42.2}{\degree}$), \textit{i.e.} within $b$-$c$ plane. Magnetic structure is formed from an admixture of two irreps: $m\Sigma_{3}$ and $m\Sigma_{4}$. \textsc{fullprof} fitting statistics: $R_{wp}=23.2\%$ and $\chi^{2}=0.7981$}\label{fig:si-f}
  \end{subfigure}

  \caption{Rietveld refinements of trial magnetic structures for Phase 4, with $T=\qty{40}{\kelvin}$. Since the powder neutron diffraction data was taken in proximity to $T_{N}$, the magnetic moment and signal are weak. Therefore, we chose to analyze the data in the medium resolution bank ($2\theta\sim\num{90}\si{\degree}$) where the signal-to-noise ratio is optimized due to the radial collimator geometry~\cite{chapon2011wish}. This data set has $R_{exp}=25.9\%$~\cite{rodriguez1993recent}. Note that powder neutron diffraction cannot distinguish between single-Q and multi-Q structures and that it is insensitive to phase differences between irrep components for multi-irrep models; \textit{e.g.} helical and SDW models are indistinguishable. In this context, we applied a single-Q SDW model with a propagation vector of $(0.224,0,0)$ for the trial solutions.}
  \label{fig:si-grid}
\end{figure*}

\newpage
\section{Generalized Basis Solutions of the Magnetic Ground State}

\begin{figure}[h]
\centering
\includegraphics[width=1\linewidth]{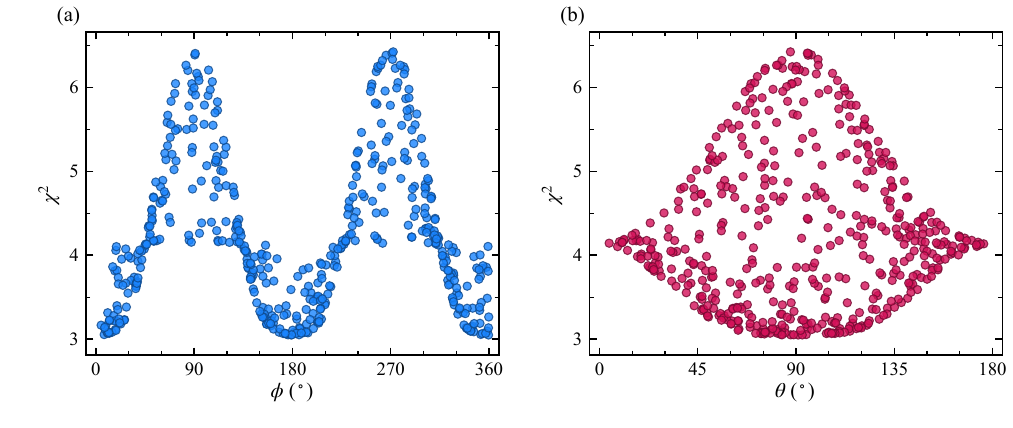}
\caption{Results from 500 neutron powder Rietveld refinements of the double-Q ground state solution in \ce{GdRu2Si2} with randomly selected orthonormal bases for the moment directions. We parametrize the moment direction of the $\mathbf{q}_{2}$ component of the magnetic structure with the angles $\phi$, the angle of the moment direction within the $a$-$b$ plane measured from $a$, and $\theta$, the angle between the moment and the $c$-axis (a) shows the $\chi^2$ value of the refinements versus the $\phi$ angle of the moment-axis. (b) shows the $\chi^2$ value of the refinements versus the $\theta$ angle of the moment-axis.}

\label{fig:OrthoBasis}
\end{figure}

The general moment-direction SDW of Phase 4 raises the possibility that the double-Q ground state could also form with a low-symmetry basis. Note that here we are referring to the basis of the magnetic moments, and not the basis of the propagation vectors: these are firmly established in the crystallographic basis with $\mathbf{q}_{1}=
(0.22,0,0)$ and $\mathbf{q}_{2}=(0,0.2246,0)$~\cite{khanh2022zoology,wood2023double}. The ansatz for the ground state is given by,

\begin{equation}
    \mathbf{m_{r}}=\begin{bmatrix} 0 \\m_{1}\cos(k_{1})\\m_{1}\sin(k_{1})\end{bmatrix} + \begin{bmatrix}m_{2}\sin(k_{2}) \\0 \\0 \end{bmatrix} + \begin{bmatrix}0 \\m_{3}\cos(k_{1}+2k_{2}) \\m_{3}\sin(k_{1}+2k_{2})\end{bmatrix},
\label{solution}
\end{equation}

\noindent where $k_{1,2}=-2\pi\mathbf{q}_{1,2}\cdot\mathbf{r}$ and $m_{1,2,3}$ are the amplitudes of the moments associated with each component of the magnetic structure. The ansatz has a constant moment ($\left| \mathbf{m_{r}}\right|$) provided $4m_{1}m_{3}=m^{2}_{2}$~\cite{wood2023double}. Crucially, the ansatz is valid for an arbitrary choice of orthonormal basis. In our previous study, we only explored a limited range of high-symmetry orthonormal bases; however, in view of the Phase 4 magnetic structure determination, we have trialed 500 randomly chosen orthonormal bases on the zero-field ground state powder measurements ($T = \qty{1.5}{\kelvin}$) presented in Wood \textit{et al.,}~\cite{wood2023double}. This could not be implemented natively in the \textsc{fullprof} refinement software. Therefore, a wrapper around the \textsc{fullprof} input file was implemented to randomly select an orthonormal basis on the surface of a unit sphere. The moment directions in the input file were fixed, and the moment sizes $m_{1}$ and $m_{2}$  were the two free parameters in the fitting. As in the original ground state study, insufficient counting statistics for the $\mathbf{q}_{1}+2\mathbf{q}_{2}$ peaks due to background and absorption in the powder data precluded fitting $m_{3}$ independently. The orthonormal basis for the magnetic structure fitting is defined by the SDW moment direction ($\hat{\mathbf{e}}(\phi,\theta)$) on $\mathbf{q}_{2}$, since the helical moment components on $\mathbf{q}_{1}$ sit in the plane perpendicular to $\hat{\mathbf{e}}(\phi,\theta)$ and powder neutron diffraction is insensitive to the relative phase difference between $\mathbf{q}_{1}$ and $\mathbf{q}_{2}$. We therefore plot the $\chi^2$ value of the refinements versus $\phi$, the angle of the moment direction within the $a$-$b$ plane measured from the $a$-axis, and $\theta$, the angle between the moment and the $c$-axis, in Fig.~\ref{fig:OrthoBasis}. Clearly $\chi^2$ is minimized with $\theta=90^\circ$ and $\phi=0^\circ$ or $180^\circ$ which corresponds to moments in the $a$-axis crystallographic direction, which means the SDW propagating on $\mathbf{q}_{2}$ adopts the symmetry of the $m\Sigma_{3}$ irrep. The optimal refinement obtained essentially identical fitting statistics and parameters to the originally proposed ground state solution with a $b$-$c$ plane helix propagating on $\mathbf{q}_{1}$ and the orthogonal spin density wave on $\mathbf{q}_{2}$~\cite{wood2023double}.

\FloatBarrier
\bibliography{refs}
\bibliographystyle{apsrev4-2}